\documentclass[runningheads]{llncs}
\usepackage{graphicx}

\usepackage{xcolor}
\usepackage{color}
\usepackage{colortbl}
\usepackage{url}
\usepackage{subfig}
\usepackage[export]{adjustbox}
\usepackage{hyperref}
\colorlet{punct}{red!60!black}
\definecolor{background}{HTML}{EEEEEE}
\definecolor{delim}{RGB}{20,105,176}
\colorlet{numb}{magenta!60!black}

\usepackage{listings}

\lstdefinestyle{mine}{
  language=java,
   basicstyle=\fontsize{9}{11}\ttfamily,
    numberstyle=\scriptsize,
  numbers=none,
  stepnumber=1,
  numbersep=8pt,
  tabsize=3,
  showspaces=false,
  showstringspaces=false,
      frame=lines
    backgroundcolor=\color{background}
}

\begin{document}
\title{Increasing the Reusability of Enforcers with Lifecycle Events\thanks{This work has been partially supported by the H2020 Learn project, which has been funded under the ERC Consolidator Grant 2014 program (ERC Grant Agreement n. 646867) and the GAUSS national research project, which has been funded by the MIUR under the PRIN 2015 program (Contract 2015KWREMX).}}
%
%
\author{Oliviero Riganelli \and
Daniela Micucci \and
Leonardo Mariani}
\authorrunning{O. Riganelli et al.}
%
\institute{University of Milano Bicocca, Viale Sarca 336, 20126 Milan, Italy
\email{\{riganelli,micucci,mariani\}@disco.unimib.it}}
\maketitle              
\begin{abstract}
Runtime enforcement can be effectively used to improve the reliability of software applications. However, it often requires the definition of ad hoc policies and enforcement strategies, which might be expensive to identify and implement. This paper discusses how to exploit lifecycle events to obtain useful enforcement strategies that can be easily reused across applications, thus reducing the cost of adoption of the runtime enforcement technology. The paper finally sketches how this idea can be used to define libraries that can automatically overcome problems related to applications misusing them.  
\end{abstract}
\section{Introduction} \label{sec:introduction}

Runtime enforcement techniques are effective solutions for guaranteeing that software applications satisfy certain correctness policies at runtime~\cite{Ligatti:2009:REN}. When using runtime enforcement, developers are typically in charge of identifying the policies that must be enforced, defining a strategy to enforce them, and finally implementing the software enforcer that applies the strategy. 

The enforced policies are often \emph{application-specific}, that is, policies are defined ad hoc for the target application. Working with application-specific policies might be quite expensive. In fact every time a new application is considered, new policies must be identified, and the modelling and implementation activities must be repeated from scratch.

Interestingly policies may also refer to libraries and components that can be reused across applications being themselves eligible for reuse. \emph{Reusable policies}  are extremely important because they can alleviate the developers from the burden of identifying both the policies to be enforced and the corresponding enforcement strategies. Developers could simply reuse policies and enforcement strategies while they reuse libraries, 
de facto simplifying the application of runtime enforcement techniques. 

Unfortunately, the definition of reusable policies and enforcement strategies can be challenging. Since the context of use of a library is not known a priori, a reusable policy and the corresponding enforcement strategy could be defined referring to the operations of the library only. For example, a reusable policy of a library for interacting with the file system may require that a file is opened before any content is written in the file. However, several relevant policies may depend not only on the usage of a  library, but also on the behavior of the application that interacts with the library. For instance, a policy that forces an app to close a file before its execution is suspended depends on both the library and the app, and cannot be specified referring to the library only.

There is a popular class of software applications that naturally facilitate both the identification of reusable policies and the definition of enforcement strategies. 
We call them \emph{life-cycle based applications}. They are applications whose units of composition are modules with an explicitly documented life-cycle model. There is a huge number of life-cycle based applications. For example, Android apps are composed of activities with  a known life-cycle model and with callbacks that are invoked when there is a change in the state of the app; similarly Spring applications are composed of components with a known life-cycle model and callbacks. The same applies to many other contexts, such as Web applications, multi-threaded applications, and so on.  

Life-cycle based applications have the important advantage of responding to the same life-cycle and implementing the same callbacks, regardless of what a specific application does. Thus policies and enforcement strategies can exploit this information to consider some aspects of the behavior of the application, still remaining reusable. We call these reusable policies \emph{life-cycle based policies} and the corresponding strategies \emph{life-cycle based enforcement strategies}. 

We further elaborate the concept of life-cycle based application and policy in Section~\ref{sec:enforcement}. We show how we exploited these concepts to define \emph{proactive libraries}, a class of libraries augmented with reusable enforcement strategies, in Section~\ref{sec:seams}. We provide final remarks in Section~\ref{sec:conclusions}.

\section{Life-cycle Based Policies} \label{sec:enforcement}

The life-cycle of a software unit specifies the possible states of the unit and the events that can cause the transition between two states. Units with a non-trivial and well-defined lifecycle are typically executed and managed by a framework that explicitly controls their life, invoking callback methods when there is a state transition. For example, Android activities have callback methods that are invoked when an application is started and suspended. Similarly, Web components have callback methods that are invoked when they are created and destroyed. 

These callback methods are pervasively present in life-cycle based applications. For instance, every  activity in every Android application implements the same callback methods. This is an important aspect that eases the definition of both reusable policies and reusable enforcement strategies that can be generally valid for every application of a specific domain. For instance, a policy about an Android library can also refer to callback methods without any loss of generality.

Policies with life-cycle events are particularly relevant. Applications may have to implement non-trivial behaviors in reaction to state transitions~\cite{Felix_2018,React_2018,Kubernetes_2018,Android:Lifecycle:website,OSGI_2018,Spring_2018}, and 
this may lead to 
faulty applications, for instance applications with faulty library interactions~\cite{Hou:API:2011,Wang:API:2013}. 

Although these policies might be non-trivial to address, they are easy to find in the documentation of libraries and systems and can be the basis for the design of reusable policies. We report below three examples of reusable policies that can be defined for completely different life-cycle based systems. 

The \texttt{onPause()} method is an Android callback that is automatically executed when a user stops interacting with an activity and is relevant to several correctness policies. For instance, 
an activity that is paused after acquiring the \texttt{Camera} must release it otherwise the camera might be unusable from other activities\footnote{\url{https://developer.android.com/guide/topics/media/camera\#release-camera}}.

In the OSGi Java framework~\cite{OSGI}, application bundles can be started, stopped, installed, and uninstalled remotely without rebooting. 
The execution of these operations must obey to specific policies. For example, stopping a bundle requires unregistering every previously registered service~\cite{OSGI_2018}.

React is a JavaScript library widely used to build encapsulated components that can be composed to create complex Web UIs \cite{React}. Each component has several life-cycle callback methods that can be overridden to execute custom code at particular times in the component's life-cycle. For example, the method \texttt{componentWillUnmount()} is invoked immediately before a component is unmounted and destroyed. The library documentation requires applications to implement specific operations when this callback is executed, such as invalidating timers, deleting network requests, or cleaning up subscriptions
~\cite{React_2018}.

Note that all these examples are cases of policies that can be arbitrarily reused across applications since they exploit information about life-cycle events and library APIs. 
These policies would be impossible to define without exploiting the information about life-cycle events. 

In the next section, we show how we exploited this concept to define \emph{proactive libraries}, that is, libraries equipped with life-cycle based enforcement strategies 
We present proactive libraries in the  Android domain because it is the most popular among the application domains described above, and because it has been already used as application domain in related work~\cite{Riganelli:RV:2017,Falcone:Runtime:2012,Riganelli:HealingDataLos:IWSF:2016}.

\section{Proactive Libraries} \label{sec:seams}
Let us refer to the \textit{Plumeria}\footnote{\url{https://github.com/DonLiangGit/Plumeria}} app, a simple Android app, to illustrate the concept of \emph{proactive library}~\cite{Riganelli:ProactiveLibraries:SEAMS:2017}. \textit{Plumeria} has a fault, that is, one of its activities does not release the camera when it is suspended, as a consequence the camera becomes inaccessible to the other apps of the device. 
This is a classic resource leak problem that could be avoided by enforcing the policy presented in Section 2. In particular, if the camera API is released as a proactive library, this problem would never show up because it would be automatically detected and fixed by the enforcement mechanism embedded in the proactive library.


Proactive libraries are standard libraries augmented with the built-in capability of enforcing reusable policies at runtime. 

\begin{figure}[ht!]
\begin{center}
\vspace{-0.5cm}
  \includegraphics[width=0.77\textwidth]{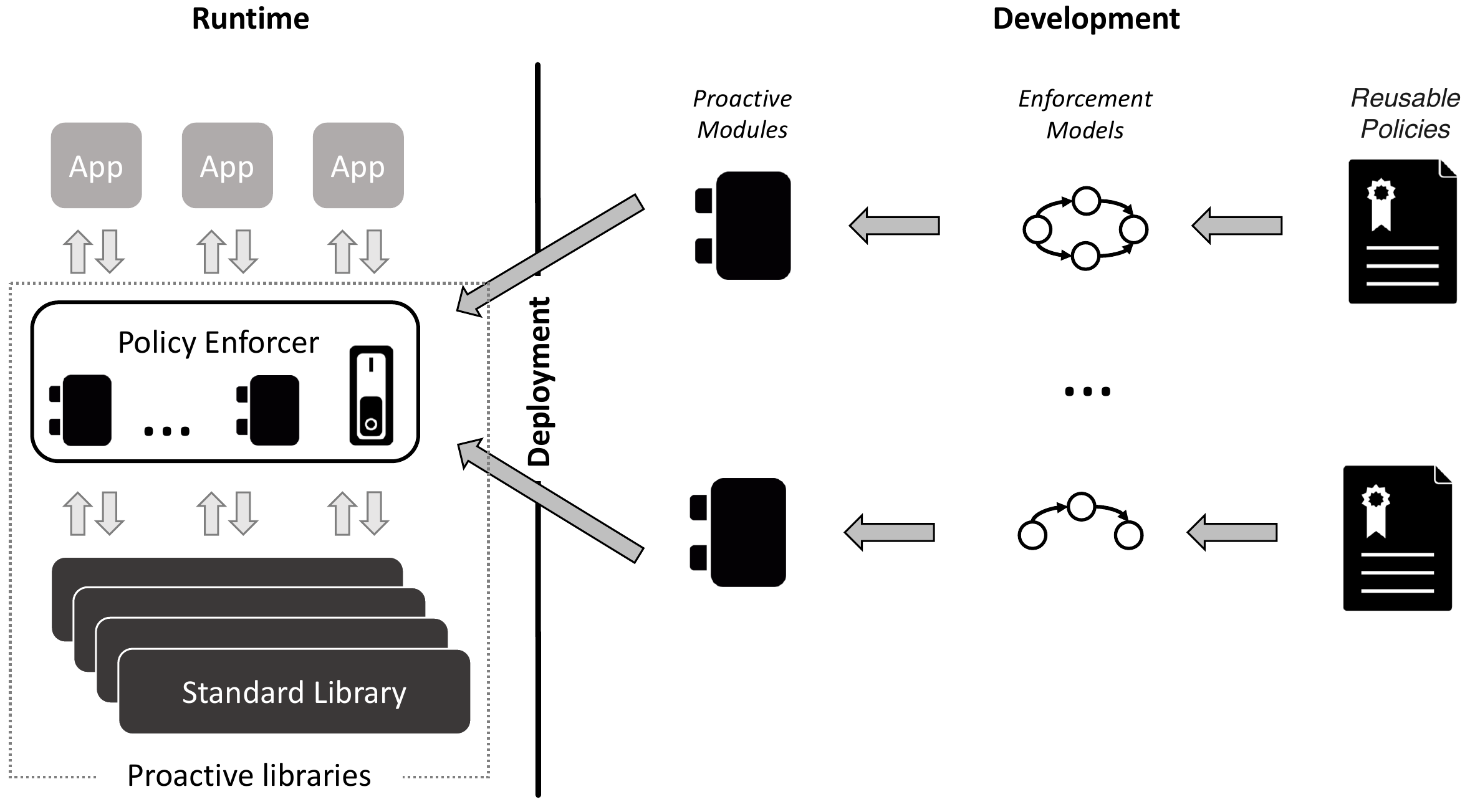}
\caption{The generation process of proactive libraries.}
\label{fig:proactivelibrary}
\vspace{-1cm}
\end{center}
\end{figure}

Figure~\ref{fig:proactivelibrary} shows the generation process of proactive libraries. 
%
%
%
The process distinguishes the development and the runtime phases. 

At development time, developers start from the identification of reusable correctness policies, that is, natural language statements that specify how the application should use a library according to the status of both the application, detected through the execution of its life-cycle callback methods, and the library, detected through the execution of its API methods. 
The reusable correctness policy that ensures the correct usage of the camera is: ``\emph{An activity that is paused while having the control of the camera must first release the camera}.''

Correctness policies are used to derive enforcement models that define how to react to correctness policies violations. 
We use edit automata~\cite{Ligatti:Edit:JIS:2005} to define the enforcement models because they naturally support the definition of enforcement rules by means of events to be intercepted, inserted and suppressed, and they could be also verified~\cite{Riganelli:RV:2017}. 
The definition of an enforcement model does not require any knowledge about the app that uses the API, but it uniquely requires the knowledge of the API and of the Android callback methods, which are the same for any app. 

\begin{figure}[ht!]
\begin{center}
\vspace{-0.4cm}
  \includegraphics[width=6cm]{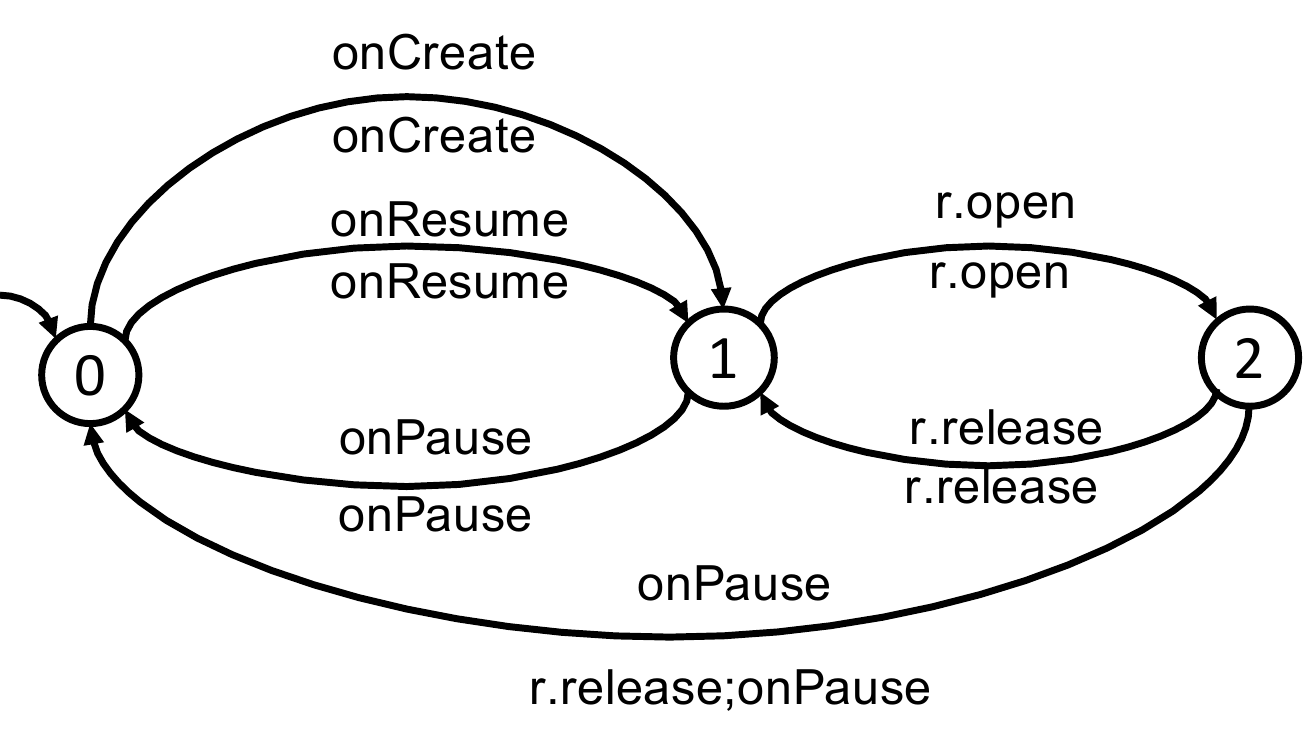}
\caption{Simplified enforcement model for the \texttt{Camera}.}
\label{fig:simplifiedmodel}
\vspace{-1cm}
\end{center}
\end{figure}

Figure~\ref{fig:simplifiedmodel} shows a slightly simplified enforcement model that forces the release of the \texttt{Camera} when the activity is paused without releasing the \texttt{Camera}. The prefix \texttt{r} is used to distinguish the calls to the API methods from callbacks. To keep the example real but small, the enforcement model does not include the part that reassigns the \texttt{Camera} to the activity once its execution is resumed. 

To actually enforce the policy in the target environment, the enforcement models are turned into \emph{proactive software modules} that intercept the execution of life-cycle callback methods and API methods, and produce additional invocations when needed, according to the enforcement model.   

 
Since proactive modules are activated by the invocation of specific methods, their execution in the user environment is controlled by a \emph{policy enforcer} that intercepts the events and dispatches them to the deployed proactive modules. The policy enforcer also controls the activation and deactivation of the proactive modules, which can be turned off and on by the user. 

The language and frameworks to implement the proactive modules and the policy enforcer depend on the target environment. In the case of Android, we use the Java Xposed framework~\cite{Xposed_2016}, which allows to cost-efficiently intercept method invocations and change the behavior of an Android app using run-time hooking and code injection mechanisms. 

In our experience, we successfully used proactive libraries to automatically overcome several problems present in Android apps~\cite{Riganelli:ProactiveLibraries:SEAMS:2017}.

\section{Conclusions} \label{sec:conclusions}
Research on runtime enforcement has already delivered both\linebreak theoretical~\cite{Ligatti:Edit:JIS:2005,Bielova2011,Falcone2012,Ligatti:2009:REN}
 and practical results \cite{Kumar:REDB:WISE:2015,Falcone:Runtime:2012,Halle:2010:REWEB,Riganelli:ProactiveLibraries:SEAMS:2017}. However, identifying policies, specifying enforcement strategies, and implementing the corresponding enforcers is still a difficult and time consuming task. Reusable policies, as discussed in this paper, can relieve developers from this tedious and error-prone task, facilitating reuse and easing the practical adoption of the runtime enforcement technology.


We plan to extend our work on runtime enforcement in three directions. \emph{Automatic code generation of runtime enforcement mechanisms}: since manually implementing runtime enforcement mechanisms is particularly difficult and expensive, we plan to define a model-driven software development process and the corresponding tool chain to automatically derive enforcer code from the models. \emph{Automatic testing of software enforcers}: To achieve highly reliable and safe enforcing mechanisms, we need techniques specifically defined to validate the behavior of software enforcers, which have the distinguishing characteristic of being designed to dynamically change the behavior of other software applications, causing hard to predict side effects. \emph{Public repository of software enforcers}: Since life-cycle based enforcement strategies are application-independent, publishing well developed software enforcers in a public repository is important to facilitate the distribution of plug-and-play enforcement strategies that can be easily exploited by developers.

\bibliographystyle{splncs04}

\end{document}